\documentstyle[twoside,fleqn,espcrc2]{article}
\def\lsim{\mathrel{\rlap{\lower4pt\hbox{\hskip1pt$\sim$}}
    \raise1pt\hbox{$<$}}}         
\def\gsim{\mathrel{\rlap{\lower4pt\hbox{\hskip1pt$\sim$}}
    \raise1pt\hbox{$>$}}}         

\title{Uncertainties in the Solar Neutrino Flux}
\author{W.C. Haxton\address{Institute for Nuclear Theory, Box 351550, and
Department of Physics, Box 351560\\
University of Washington, Seattle, WA 98195, USA}}

\begin{document}

\input psfig.sty

\begin{abstract}
I discuss three issues relevant to solar neutrino flux measurements:  cross 
section uncertainties in pp chain reactions, uncertainties in the GALLEX/SAGE 
response to $^7$Be and $^{51}$Cr neutrinos, and the implications of 
helioseismology for nonstandard suns with mixed cores.  A few comments are 
also offered on $\nu_e \leftrightarrow \nu_\tau$ oscillations, cosmologically 
interesting neutrino masses, and recent proposals for supernova neutrino 
observatories.
\end{abstract}
\maketitle

\section{INTRODUCTION}

It is a pleasure to be present for this historic meeting hosted by the 
Superkamiokande collaboration.  In this talk I will address three issues 
affecting the solar neutrino flux and one connected with future detectors for 
supernova neutrinos.

\subsection{Nuclear Physics of the pp Chain}

One of the crucial inputs into the solar model is the network of nuclear 
reactions 
comprising the pp chain (and CNO cycle).  This network involves nonresonant 
charged particle reactions occurring at center-of-mass energies well below the 
height of the Coulomb barrier.  As the solar core temperature $T_c \sim 1.5 
\cdot 10^7 K$, the typical kinetic energy for a nucleus in the core is 
$<E> \sim$ 2 keV.  The competition 
between the Coulomb barrier and Boltzman distribution leads to a typical 
energy for reacting nuclei of 
$\langle E_{\mathrm {reacting}}\rangle \sim$ 10 keV, a value that is 
generally lower than that where such reactions can be measured in the 
laboratory.  Thus the task for nuclear physicists is to measure such reactions 
as accurately as possible over the accessible range of laboratory energies, 
then extrapolate these measurements to the energies relevant for the sun.

In the case of the driving reaction of the pp chain
\begin{equation}
p + p \to \, ^2\mathrm{H} + e^+ + \nu_e\;,
\end{equation}
the cross section is not measurable in the laboratory.  Thus we must rely on 
theory.  Fortunately deuterium is the simplest nucleus, and its properties are 
very well reproduced by NN potentials (Bonn, Paris, Argonne v18, etc.) 
carefully fit to phase shifts.  The calculated cross section depends on the
 accuracy with which the 
axial vector coupling $g_A$ is known and on two-body corrections to the 
space-like component of the axial current, which are fortunately of order 
(v/c)$^2 \sim$ 1 \%, where v is a typical bound nucleon velocity.

The other major reactions of the pp chain are measureable, but generally not at the 
low energies relevant to our sun.  The necessary extrapolation of the cross 
section $\sigma (E)$ to lower energies is accomplished via the S factor
\begin{equation}
\sigma(E) \equiv {S(E) \over E} e^{-2 \pi Z_1 Z_2 \alpha/\beta} 
\end{equation}
where $E$ is the center-of-mass energy, $Z_1$ and $Z_2$ are the charges of the 
interacting nuclei, and $\beta$ is the relative velocity.  The introduction 
of $S(E)$ removes the s-wave Coulomb interaction of point particles and thus 
provides a much smoother quantity for use in extrapolating data.  $S(E)$ 
depends on a number of physical effects - nuclear finite size, atomic 
screening corrections, higher particle waves, etc. - that the theorist must 
evaluate before this extrapolation can be done.

Because the solar neutrino problem is at a crucial juncture, a group of about 
40 experts recently met at the Institute for Nuclear Theory, Seattle, to 
discuss the nuclear physics of the pp chain and CNO cycle.  The questions 
addressed included the best current values for cross sections, critiques 
 of assigned uncertainties, and recommendations for future 
experimental and theoretical work that could further improve our understanding 
of the nuclear physics.  The summary of this workshop will appear in Reviews
 of Modern Physics 
(October, 1998) and is also available on the LANL preprint archive [1].

\begin{figure}[htb]
\psfig{bbllx=1.8cm,bblly=10cm,bburx=17cm,bbury=21cm,figure=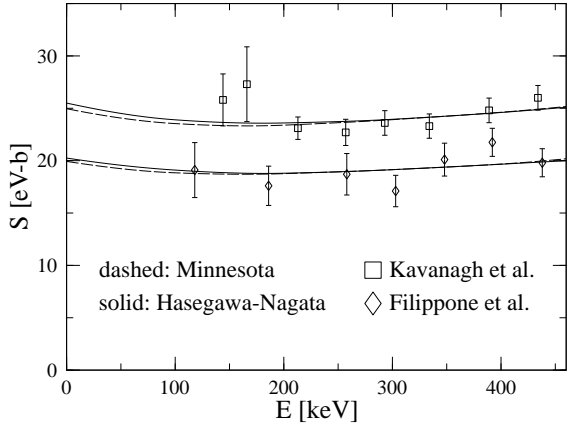,height=1.92in}
\caption{The $^7$Be($p,\gamma)^8$B S-factor as measured by Filippone et al.
[3] and by Kavanagh et al. [4].  For each data set two theoretical
extrapolations to S(0), reflecting different choices for the strong potential,
are shown [2].}
\label{fig:one}
\end{figure}

While I cannot give an adequate summary here, I will mention two of the 
reactions where significant changes were recommended.  The first of these is 
$^7$Be(p, $\gamma)\, ^8$B, where the standard S$_{17}$(0)$ \sim$ 22.4 eVb is 
that given by Johnson et al [2].  Measurements of S$_{17}$(E) are complicated 
by 
the need to use radioactive targets and thus to determine the areal density of 
the $^7$Be target nuclei.  Two techniques have been employed, measuring the 
rate of 478 keV photons from $^7$Be decay or counting the daughter $^7$Li 
nuclei via the reaction $^7$Li (d,p)$^8$Li.  The low-energy data sets [3,4] 
for 
S$_{17}$(E) disagree by 25\%, a systematic effect apparent in Fig. 1.  Each 
data set is consistent with theory in its dependence on E:  this dependence is 
simple in the illustrated low energy region as it is determined by the 
asymptotic nuclear wave function. 

The Seattle working group on S$_{17}$(E) found that only one low-energy data 
set, that of Filippone et al. [3], was described in the published literature in 
sufficient detail to be evaluated.  The target activity in that experiment had 
been measured by both 478 keV gamma rays and by the (d,p) reaction, with 
consistent results.  The resulting recommended value was thus based on this 
measurement, yielding
\begin{equation}
S_{17} (0) = 19^{+4}_{-2} eV \; b\;, \; 1 \sigma \;.
\end{equation}
Since the workshop, two developments have occurred.  The 
Orsay/Bordeaux/Paris-Sud/ USTHB group published [5] a new measurement of
 S$_{17}$ 
(0) = 18.5 $\pm$ 1.0 eV b, while a preliminary value from the 
Weizmann/Troitzk/Mainz/Isolde group of S$_{17}(E_p$ = 1.2 MeV) = 22.5 $\pm$ 
2.5 eV b has been announced [6].

The $^3$He($\alpha,\gamma) ^7$Be reaction has been measured by two techniques, 
by counting the capture $\gamma$ rays and by detecting the resulting $^7$Be 
activity.  While the two techniques have been used by several groups and have 
yielded separately consistent results, the capture $\gamma$ ray value 
S$_{17}$(0) = 0.507 $\pm $ 0.016 keV b is not in good agreement with the $^7$Be 
activity value 0.572 $\pm$ 0.026 keV-b.  The Seattle working group concluded 
that the evidence for a systematic discrepancy of unknown origin was 
reasonably strong and recommended that standard procedures be used in 
assigning a suitably expanded error.  The recommended value S$_{34}$ (0) is 
0.53 $\pm$ 0.05.

These and other recommended values were recently incorporated 
into the 
Bahcall and Pinnsoneault (BP98) solar model calculation [7].  While the 
workshop's 
recommended values involve no qualitative changes, there is some broadening of 
error bars and a downward shift in S$_{17}$(0), leading to the lower BP98 
$^8$B flux.  The workshop's Reviews of Modern Physics article summarizes a 
substantial amount of work on topics not discussed here:  screening effects, 
weak radiative corrections to and exchange current effects on p+p, the atomic 
physics of $^7$Be + e$^-$, etc.  Much of this discussion was useful in 
evaluating possible uncertainties in solar microphysics, and in identifying
opportunities for reducing those uncertainities.

\subsection{The Nuclear Physics of the \\ 
GALLEX/SAGE $^{51}$Cr Calibrations}

The $^{51}$Cr neutrino source experiments provide an important check on the 
overall gallium detector operations under few atom, hot chemistry conditions.  
The issue discussed here, and which was mentioned in the earlier experimental 
talks, is the potential complication due to contributions of uncertain 
strength to the $5/2^-$ and $3/2^-$ $^{71}$Ge excited states (see Fig. 2).
\begin{figure}[h]
\psfig{bbllx=2cm,bblly=5.5cm,bburx=18cm,bbury=22cm,figure=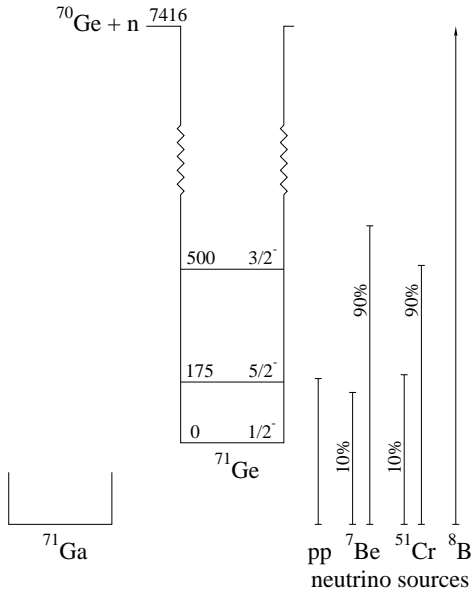,height=3.0in}
\caption{Level scheme for $^{71}$Ge showing the excited states that contribute 
to absorption of pp, $^7$Be, $^{51}$Cr, and $^8$B neutrinos.}
\end{figure}

The results of the source experiments can be normalized to the known $^{71}$Ge 
ground state contribution [8], yielding
\begin{eqnarray*}
R_o = E \left[1 + 0.67 {{\mathrm {BGT}} (5/2^-) \over {\mathrm {BGT}} 
({\mathrm {gs}})} + 
0.22 {{\mathrm {BGT}} (3/2^-) \over 
{\mathrm {BGT}} ({\mathrm {gs}})} \right]
\end{eqnarray*}
\begin{equation}
= \left\{ \begin{array}{ll}
0.98 \pm 0.08,&{\mathrm {GALLEX}} ~~[9]\\
1.00 \pm 0.13,&{\mathrm {SAGE}} ~~[10]
\end{array}
\right.
\end{equation}
where $E$ represents any departure of the efficiency from the value determined 
from tracers and used by the experimentalists in their analyses.  The 
dependence of the results on the unknown BGT values is explicit.  Clearly the 
conclusion $E \sim 1$ requires an independent determination that the unknown 
BGT values are much smaller than the ground state value BGT(gs).

It had been assumed that forward-angle (p,n) charge exchange measurements 
determine the unknown BGT values
\begin{eqnarray*}
{{\mathrm {BGT}}(5/2^-) \over {\mathrm {BGT}} ({\mathrm {gs}})} \lsim 0.06 
\end{eqnarray*}
\begin{equation}
{{\mathrm {BGT}} (3/2^-) \over {\mathrm {BGT}}  ({\mathrm {gs}})} 
 = 0.12 \pm 0.02 
\end{equation}
But extensive investigations [11,12] of the proportionality between (p,n) cross 
sections and known weak interaction BGT values have shown that the 
relationship is a complicated one.  Existing discrepancies can be removed by 
the assumption that forward-angle (p,n) effective operator 
contains a spin-tensor contribution of relative strength 
$\delta \sim$ 0.1, in addition to the Gamow-Teller operator,
\begin{eqnarray*}
\hat O_{(p,n)} =  \sigma (i) \tau_+ (i) + 
\end{eqnarray*}
\begin{equation}
\delta \sqrt{8 \pi} [Y_2 
(\Omega_i) 
\otimes \sigma (i)]_{J = 1} \tau_+ (i). 
\end{equation}
Simple considerations of the nuclear structure of $^{71}$Ga and $^{71}$Ge 
suggest that the tensor operator might be particularly troublesome for the 
$^{71}$Ga(3/2$^-) \to ^{71}$Ge(5/2$^-$) transition to the first excited state.  
The naive description of this transition is
\begingroup
     \def\theequation{7}
     \begin{equation}
     1f_{5/2} (n) \to 2p_{3/2} (p),
     \end{equation}
     \endgroup
\noindent
an $\ell$-forbidden amplitude that generates an enormous spin-tensor and
vanishing Gamow-Teller contributions.  Well-known transitions of a similar 
character, e.g. to the first excited state in $^{39}$K(p,n) $^{39}$Ca, have 
produced discrepancies between (p,n) and weak interaction transition 
probabilities of factors of $\sim$100.  The conclusion [8] is that it might be unwise to use 
the $^{71}$Ga (p,n) results as a reliable independent measurement of BGT 
(5/2$^-$).

To explore this further, I did a large-basis shell model calculation [8] of 
the 
$^{71}$Ga $\to ^{71}$Ge weak and (p,n) transitions.  The results agree 
reasonably with what is known experimentally:  the calculated 
BGT (g.s.) = 0.051, compared to the experimental value 0.087, while the 
calculated (p,n) BGT (5/2$^-$), corresponding to the operator in Eq. (6), is 
0.0006, in agreement with the experimental bound $<$0.005.  
However the latter result stemmed from a cancellation between the GT and 
spin-tensor operators comprising O$_{\mathrm {(p,n)}}$,

\begin{equation}
\langle 5/2^- || O_{p,n} || 3/2^- \rangle = 0.264 - \delta 2.23.
\end{equation}

The cancellation between the second term - an enormous spin-tensor amplitude - 
and the GT amplitude leads to a much larger beta decay BGT(5/2$^-$) than 
would be 
allowed by in a more naive interpretation of the (p,n) BGT value.  Thus this 
is an explicit demonstration that previous bounds on BGT (5/2$^-$) are too 
aggressive.

It is argued in Ref. [8] that, by relying on the shell model calculation of 
the very strong tensor amplitude in Eq. (8), a reasonable range of beta decay  
BGT (5/2$^-)$ can be extracted from the (p,n) measurements.  This results in a 
corresponding change in the $^{51}$Cr cross section from the previous standard 
value
\begingroup
     \def\theequation{9a}
     \begin{equation}
\sigma(^{51}{\mathrm {Cr}}) = (5.81^{+0.21}_{-0.16}) \cdot 10^{-45} {\mathrm
{cm}}^2     
     \end{equation}
     \endgroup
\noindent
to
\begingroup
     \def\theequation{9b}
     \begin{equation}
\sigma (^{51}{\mathrm {Cr}}) = (6.39 \pm 0.68) \cdot 10^{-45} {\mathrm 
{cm}}^2
     \end{equation}
     \endgroup
\noindent
where the error in Eq. (9b) represents that due to excited state uncertainties 
only.  If this value is used in Eq. (4), one finds
\begingroup
     \def\theequation{10}
     \begin{equation}
E =\left\{ \begin{array}{ll}
0.86 \pm 0.07 \pm 0.09,&{\mathrm {GALLEX}} \\
0.875 \pm 0.11 \pm 0.09,&{\mathrm {SAGE}}
\end{array}
\right.
     \end{equation}
     \endgroup
\noindent
where the first uncertainty in the source experiment error while the second 
corresponds to the $^{51}$Cr cross section.  Note that E $\sim$ 1 is allowed, 
though it is certainly not demanded.  It is important to note that the 
difference between (9a) and (9b) is one of an extended error range:  all of 
the range in (9a) that is  attributable to excited state uncertainties is 
allowed in (9b). It is also notable that the cross section uncertainty in Eq. 
(10) is comparable to the source experiment uncertainty.

\begin{figure}[h]
\psfig{bbllx=1.6cm,bblly=4.5cm,bburx=12cm,bbury=24.3cm,figure=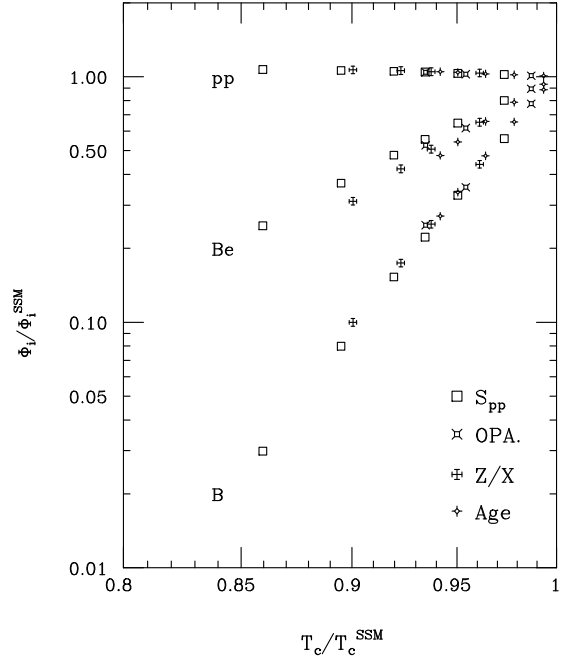,height=3.4in}
\caption{The response of the pp, Be, and B neutrino fluxes to the indicated
variations in solar model input parameters, displayed as a function of the
resulting central temperature T$_c$.  (From Castellani et al. [13].)}
\end{figure}

The conclusion is that the source experiments have become $^7$Be neutrino 
cross section measurements.  Indeed, one can express the GALLEX/SAGE responses 
to the $^7$Be neutrinos in terms of $R_o$, independent of almost all nuclear 
physics uncertainties.  This means of course that other tests of E$\sim$ 1 
under few-atom, hot chemistry conditions take on added importance.  Thus the 
GALLEX $^{71}$As test discussed by Prof. Kirsten at this meeting is crucial. 
The GALLEX/SAGE 
results are central to the conclusions of global analyses of the solar 
neutrino experiments that yield $\phi (^7$Be) $\lsim 0$.

\subsection{Solar Core Mixing of $^3$He and Helioseismology}
The crux of the solar neutrino problem can be captured in two experimental 
quantities.  First the $^8$B neutrino flux $\phi(^8$B), which varies 
approximately as 
$T_c^{18}$ where $T_c$ is the solar core temperature, is known to be reduced 
by about a factor of 1/3 (0.47 using the new BP98 results).  Naively this result
requires a cooler sun,
\begin{eqnarray*}
T_c \sim 0.96 T_c^{\mathrm {SSM}},
\end{eqnarray*}
where the superscript SSM denotes the standard solar model result.  However 
the flux ratio $\phi(^7$Be)/$\phi(^8$B), which varies as $T_c^{-10}$, also 
appears to be reduced relative to the standard model.  This then requires
\begin{eqnarray*}
T_c > T^{\mathrm {SSM}}_c
\end{eqnarray*}
with the extent of the increase depending on how strongly one wants to 
suppress this ratio.  ($\phi(^7$Be) $\sim$ 0 provides the best fit to the 
$^{37}$Cl, GALLEX/SAGE, and Superkamiokande results.)  It appears that the 
experimental results on $\phi(^8$B) and $\phi(^7$Be)/$\phi(^8$B) are thus in 
conflict, with the first requiring a cooler sun and the second a hotter one.

\begin{figure}[h]
\psfig{bbllx=0.5cm,bblly=10cm,bburx=19cm,bbury=23.3cm,figure=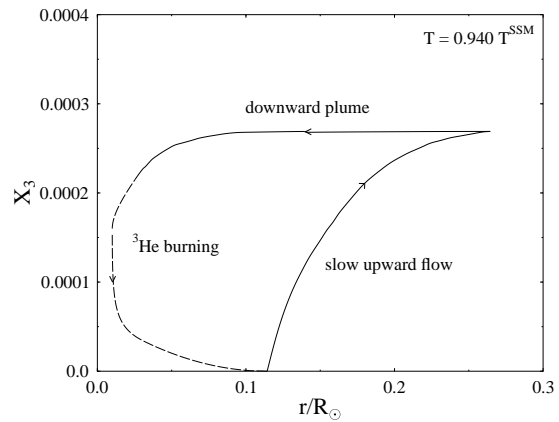,height=2.07in}
\caption{A phenomenologically derived core convection pattern that will
suppress both $\phi(^8$B) and $\phi(^7$Be)/$\phi(^8$B) [14].  The downward
flow is in plumes, rapid and localized, requiring $\sim$ few $\cdot ~10^6$
years.
 This leads to out-of-equilibrium burning of $^3$He at small r.  The slow,
broad, upward flow allows the cycle to replenish the $^3$He.  Typical upward
times are $\sim$ few $\cdot ~10^7$ years.}
\end{figure}

These arguments depend on the assumption that neutrino fluxes will track $T_c$ 
as described above, but this appears to hold remarkably well.  Figure 3, from 
Castellani et al. [13], illustrates this:  changes due to modified nuclear 
cross 
sections, opacity, lowered metallicity, and the solar age 
produce neutrino fluxes that track the resulting $T_c$ quite accurately. 
This led some in the field to argue that no nonstandard 
solar model could produce the observed pattern of neutrino fluxes.

Andrew Cumming and I decided to test this claim phenomenologically, under the 
assumption of a steady-state sun with conventional microphysics producing the 
correct luminosity.  As our procedures are described elsewhere [14], I'll just 
state here our basic result.  There appears to be only one possibility for 
constructing a steady state model with a neutrino flux pattern reasonably 
close to experiment:  the solar core must mix on timescales of $^3$He 
equilibration ($\sim$ few 10$^6$ years) in the ``elevator convection" pattern 
illustrated in Fig. 4.  This mixing produces the desired flux pattern because 
it modifies the ppI/(ppII + ppIII) and ppII/ppIII branching ratios in the 
proper way.  The former is enhanced because the resulting $^3$He enrichment 
of the 
core favors the $^3$He + $^3$He 
reaction.  The latter is reduced because the fraction of $^3$He that burns by 
$^3$He + $^4$He produces $^7$Be deep in the core, where the higher 
temperatures favor ppIII over ppII.

This exercise indicated that arguments against nonstandard solar model 
solutions based on how neutrino fluxes scale with $T_c$ are not completely general.  While the pattern 
of core mixing was derived phenomenologically, and not on physical grounds, it 
nevertheless has some physical appeal.  The possibility of core mixing 
generated by the standard solar model overstability in the $^3$He gradient was 
first discussed by Dilke and Gough.  Roxburgh discussed a persistent 
convective core as a possible consequence of the growth of the SSM $^3$He 
gradient during our sun's early convective stage.

Several astrophysical consequences of such mixing were discussed, with one, 
helioseismology, appearing problematic.  Bahcall et al. [15] and 
Fiorentini et al. [16] evaluated the helioseismological consequences of 
replacing the SSM core molecular weight profile by a constant one, such as 
would occur
for a continuously mixed model.  This yielded 8\% deviations in core sound 
speeds, far outside allowable bounds.  My summary of this work is that it 
might be viewed as an attempt to estimate the natural scale of expected 
helioseismology  changes.  However, it does not convincingly settle the 
issue because such a modification of the SSM produces a ``model" that fails to 
satisfy the equations of stellar evolution.  One can envision that it might be 
more difficult to change the sound speed profile c(r) in a dynamically 
consistent model where the pressure and density profiles are coupled through 
the condition of hydrostatic equilibrium.

A stronger argument, which arose from discussions at last December's ITP 
conference on solar neutrinos, is that 
existing helioseismology determinations of c(r) coupled with a constant molecular weight profile in the 
core would necessarily lead to an unphysical temperature profile T(r) that would 
increase away from r=0. 
It 
appears this conclusion is overstated:  adiabatically mixed models lacking 
molecular weight gradients in the core generically have profiles in T(r) that
 are convex 
downward.  [While this counterargument to the ITP discussions was originally 
made to me by Richard Epstein, 
similar remarks have been made by John Bahcall and by Doug Gough at this 
conference.]

My view is that such mixed models, while quite unlikely, are not yet definitively 
ruled out.  The issue of their viability is an important one, given the 
argument that a core mixed on the $^3$He equilibration timescale is the only 
possibility for producing acceptable neutrino fluxes in a steady state model.  
Plans are underway at Los Alamos to evolve a series of 1D models where mixing 
is included through mixing length theory, adjusting these in the usual way to 
produce the proper luminosity after 4.6 b.y. of the stellar burning. 
Helioseismology studies performed on these models should then settle the 
issue.

\subsection{Supernova $\nu_e \leftrightarrow \nu_\tau$ Oscillations and 
Cosmologically Interesting Neutrino Masses}

In a Type II supernova 99\% of the energy released by the core collapse is 
carried off by neutrinos.  The initial flavor equilibrium of neutrinos trapped 
within the core at densities $\rho \gsim 10^{12}$ g/cm$^3$, coupled with the 
flavor-dependent decoupling of neutrinos from the matter at the neutrinosphere, 
leads to an approximate equipartition of energy among the flavors and to a 
characteristic hierarchy of temperatures.  The average energy of heavy flavor 
neutrinos $\langle E_{\nu_{\mathrm {HEAVY}}} \rangle \sim $ 25 MeV, while 
$\langle E_{{\bar \nu}_e} \rangle \sim$ 16 Mev and $\langle E_{\nu_e} \rangle 
\sim$ 11 MeV.   The lower values for the $\nu_e$s and $\bar \nu_e$s reflects 
their stronger matter couplings due to charged current reactions with nucleons 
and to their greater scattering cross sections off electrons.  The lower $\nu_ 
e$ temperature, relative to $\bar \nu_e$, is due to the neutron richness of 
the matter near the neutrinosphere and resulting enhancement of $\nu_e + n \to 
p + e^-$.  The neutrino energy hierarchy $\langle E_{\nu_{\mathrm {HEAVY}}} 
\rangle > \langle E_{\bar \nu_e} \rangle > \langle E_{\nu_e} 
\rangle$ appears to be a result independent of the details
 of supernova 
modelling, in contrast to the case of solar neutrinos where fluxes depend on 
nuclear reaction networks.

\begin{figure}[h]
\psfig{bbllx=1.5cm,bblly=5cm,bburx=15.8cm,bbury=18.5cm,figure=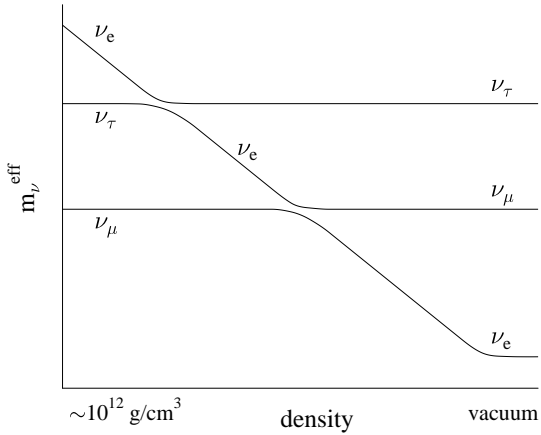,height=2.26in}
\caption{The three-flavor level crossing diagram showing two ``crossings" that
might be associated with matter enhanced oscillations of supernova neutrinos.}
\end{figure}

The spectrum of neutrinos is essentially fixed at the neutrinosphere, $\rho \sim 
10^{12}$ g/cm$^3$, a density that corresponds to a neutrino oscillation level 
crossing for $\delta m^2 \sim 10^4$ eV$^2$.  Furthermore, the density scale height at this 
density will produce an adiabatic neutrino level crossing for $\sin^2 2\theta 
\gsim 10^{-5}$.  Thus the supernova neutrino  spectrum provides a unique opportunity to 
probe neutrino oscillations.  In particular, if the neutrino masses have a 
seesaw pattern 
\begin{eqnarray*}
m_\nu \propto {m_D^2 \over M_R} \,\,\,\,\,\,\,\,\, \, m^2_D \leftrightarrow 
m^2_u : m^2_c : 
m^2_t
\end{eqnarray*}
one can fix $M_R$ according to the solar neutrino small angle solution, 
assuming the solar neutrino puzzle is due to $\nu_e \to \nu_\mu$.  Thus 
``$m_{\nu_\mu}$" $\sim$ few $\cdot 10^{-3}$ eV and ``$m_{\nu_\tau}$" $\sim$ 1 
eV, a value that would be interesting cosmologically and would induce $\nu_e 
\leftrightarrow \nu_\tau$ oscillations for supernova neutrinos.  This is 
illustrated in Fig. 5.  As a consequence of this crossing, supernova $\nu_e$s 
will emerge anomalously hot, with an $\langle E_{\nu_e}\rangle \sim$ 25 MeV 
characteristic of heavy flavor neutrinos.  An experimental demonstration that 
$\langle E_{\nu_e} \rangle > \langle E_{\bar \nu_e} \rangle$, for 
example, would provide strong evidence for oscillations and possibly provide 
information on massive tauon neutrinos.

Such an oscillation would have consequences for the supernova explosion, as it 
would enhance matter heating by the $\nu_e$s, and on nucleosynthesis, as the 
increased rate for $\nu_e + n \to e^- + p$ would drive the atmosphere above 
the protoneutron star proton rich, destroying any possibility for an r-process. 
 But the aspect on which I would like to focus is the possibility of 
distinctive oscillation signals in terrestrial supernova neutrino 
observatories.

One detector of interest, despite being primarily sensitive 
 to $\bar \nu_e$s, 
is Superkamokande.  The usual $\nu_e$ signal, elastic scattering off 
electrons, will not be altered in total rate due to a $\nu_e \leftrightarrow 
\nu_\tau$ oscillation, since this rate is proportional to the luminosity, 
which is approximately independent of flavor.  However there is some hardening 
of the spectrum of forward-scattered electrons:  the question is whether this 
is enough of a signal, given uncertainties in the supernova $\nu_e$ and 
$\nu_\mu/\bar \nu_\mu/\nu_\tau/\bar \nu_\tau$ fluxes contributing to this 
forward scattering of electrons [17].  Perhaps more interesting is the reaction 
$\nu_e + ^{16}$O $\to ^{16}$F + $e^-$, which produces a backward-peaked 
distribution of electrons that would very likely be detectable above the 
$\bar {\nu_e}$ ``background," given $\langle E_{\nu_e} \rangle \sim$ 25 MeV
 [18].  There 
has also been a suggestion that the $\gamma$-ray cascades following 
$\nu$-induced spallation reactions on $^{16}$O might provide an attractive 
signal [19].

It seems to me that a detector of a different type - one flavor specific and 
economical to operate - might be useful in supernova watches, given that the 
characteristic time between galactic events might be $\sim$ 35 years.  An 
attractive possibility is the 1 kiloton version of the iodine detector 
discussed by Lande, which would be able to view the entire galaxy with good 
statistics.  The cross section averaged over the $\nu_e$ flux is predicted to 
increase by a factor of $\sim$6.6 if there is a $\nu_e \leftrightarrow 
\nu_\tau$ oscillation.  As the luminosity constraint leads to a smaller flux of 
(undistorted) $\nu_\tau$s, this implies an increase in the iodine detector 
response of a factor of $\sim$ 2.9, a rather dramatic signal for oscillations.

Another possibility is the lead neutron spallation detector LAND proposed by 
Hargrove et al. [20].  The signal he discussed, single neutron emission, 
is not flavor specific nor is the cross section well determined.  However, 
Fuller, McLaughlin, and I [21] recently found a rather attractive 
charge-current-specific signal in this detector, the emission of multiple 
neutrons.  Due 
to details of the nuclear physics - the location of the giant resonances that 
are strongly excited by
charged and neutral current scattering - this 
channel appears to ``filter out" neutral current events, while leaving $\sim$ 
70\% of the charged current response.  Thus if multiple neutron events are 
studied, LAND becomes flavor specific and, due to the high  
threshold for reaching the giant resonances, extraordinarily sensitive to the
 $\nu_e$ temperature.  A $\nu_e 
\leftrightarrow \nu_\tau$ oscillation would produce approximately 40 times the 
number of events that would be measured in the absence of oscillations.  The 
cross section for detecting $\langle E_{\nu_e}\rangle \sim$ 25 MeV supernova 
neutrinos is a remarkable 4 $\cdot 10^{39}$ cm$^2$.  There is some work 
remaining to be done to verify these results, but the prospects appear quite 
promising.

This conference is a wonderful illustration of the power of experiments with 
astrophysical neutrinos.  Supernova neutrinos differ from atmospheric and 
solar neutrinos in that they impact the earth only once or twice a century.  
But they provide unique windows on neutrino physics, such as $\nu_\tau$ 
masses of cosmological interest.  Thus it is probably important for us to 
prepare the right complement of neutrino observatories in anticipation of the 
next galactic supernova.

This work was supported in part by the U.S. Department of Energy.


\begin{thebibliography}{99}

\bibitem{}
 E.G. Adelberger, et al., to appear in Rev. Mod. Phys.
\bibitem{}
 C.W. Johnson, E. Kolbe, S.E. Koonin, and K. Langanke, Ap. J, 
392 (1992) 320.
\bibitem{}
B.W. Filippone, A. J. Elwyn, C. N. Davids, and D. D. Koetke, Phys. Rev. Lett., 
50 (1983) 412 and Phys. Rev. C, 28 (1983) 2222.
\bibitem{}
R. W. Kavanagh, T. A. Tombrello, J. M. Mosher, and D. R. Goosman, Bull. Am. 
Phys. Soc., 14 (1969) 1209.
\bibitem{}
F. Hammache et al., Phys. Rev. Lett. 80 (1998) 928.
\bibitem{}
L. Weissman et al., Nucl. Phys. A, 630 (1998) 678; M. Hass et al., private 
communication.
\bibitem{}
J. N. Bahcall and M. H. Pinsonneault, Rev. Mod. Phys., 67 (1995) 781 and to be 
published (1998).
\bibitem{}
W. C. Haxton, Phys. Lett. B, 431 (1998) 110.
\bibitem{}
P. Anselmann, et al., Phys. Lett. B, 342 (1995) 440; W. Hampel et al., Phys. 
Lett. B, 388 (1996) 384.
\bibitem{}
D. N. Abdurashitov et al., Phys. Lett. B, 328 (1994) 234 and hep-ph/9803418 
(March, 1998).
\bibitem{}
S.M. Austin, N. Anantaraman, and W. G. Love, Phys. Rev. Lett., 73 (1994) 30; 
J.W. Watson et al., Phys. Rev. Lett., 55 (1985) 1369.
\bibitem{}
N. Hata and W. C. Haxton, Phys. Lett. B, 353, (1995) 422.
\bibitem{}
V. Castellani, S. Degl'Innocenti, G. Fiorentini, M. Lissia, and B. Ricci, 
Phys. Rev. D, 50 (1994) 4749.
\bibitem{}
A. Cumming and W. C. Haxton, Phys. Rev. Lett., 77 (1996) 4286.
\bibitem{}
J.N. Bahcall, M. H. Pinsonneault, S. Basu, and J. Christensen-Dalsgaard, Phys. 
Rev. Lett., 78 (1997) 171.
\bibitem{}
G. Fiorentini, talk presented on the ITP Workshop on Solar Neutrinos, December 
1997 (http://www.itp.ucsb.edu/online/snu/ fiorentini).
\bibitem{}
H. Minakata, private communication.
\bibitem{}
W.C. Haxton, Phys. Rev. D, 36 (1987) 2283; Y. Z. Qian and G.M. Fuller, Phys. 
Rev. D, 49 (1996) 1762.
\bibitem{}
K. Langanke, P. Vogel, and E. Kolbe, Phys Rev. Lett., 76 (1997) 2629.
\bibitem{}
C.K. Hargrove et al., Astroparticle Physics, 5 (1996) 183.
\bibitem{}
G.M. Fuller, W. C. Haxton and G. C. McLaughlin, submitted to Phys. Rev. D.



\end{thebibliography}
\end{document}